\documentclass[11pt]{article}
\usepackage{amssymb}

\usepackage{graphicx}
\usepackage{amsmath}
\usepackage{makeidx}
\usepackage{indentfirst}

\newcounter{resultnum}[section]

\newcounter{conclusionnum}[section]

\newcounter{conditionnum}[section]

\newcounter{conjecturenum}[section]

\newcounter{examplenum}[section]

\newcounter{exercisenum}[section]

\newcounter{lemmanum}[section]

\newcounter{notationnum}[section]

\newcounter{theoremnum}[section]

\newcounter{definitionnum}[section]

\newcounter{corollarynum}[section]

\newcounter{remarknum}[section]

\newcounter{propositionnum}[section]

\newcounter{acknowledgementnum}[section]

\newcounter{algorithmnum}[section]

\newcounter{axiomnum}[section]

\newcounter{casenum}[section]

\newcounter{claimnum}[section]

\newcounter{summarynum}[section]

\newcounter{problemnum}[section]

\begin{document}

\title{Finsler Black Holes Induced by Noncommutative\\
Anholonomic Distributions in Einstein Gravity}
\date{March 9, 2010}
\author{ Sergiu I. Vacaru\thanks{
sergiu.vacaru@uaic.ro, Sergiu.Vacaru@gmail.com;\newline
http://www.scribd.com/people/view/1455460-sergiu } \\
{\quad} \\
{\small {\textsl{\ Science Department, University "Al. I. Cuza" Ia\c si},} }%
\\
{\small {\textsl{\ 54 Lascar Catargi street, 700107, Ia\c si, Romania}} }}
\maketitle

\begin{abstract}
We study Finsler black holes induced from Einstein gravity as possible effects of quantum spacetime noncommutativity. Such Finsler models are defined by nonholonomic frames not on tangent bundles but on (pseudo) Riemannian manifolds being compatible with standard theories of physics. We
focus on noncommutative deformations of Schwarzschild metrics into locally anisotropic stationary ones with spherical/rotoid symmetry. There are derived the conditions when black hole configurations can be extracted from two classes of exact solutions depending on noncommutative parameters. The
first class of metrics is defined by nonholonomic deformations of the gravitational vacuum by noncommutative geometry. The second class of such solutions is induced by noncommutative matter fields and/or effective polarizations of cosmological constants.

\vskip0.1cm


\vskip3pt


\end{abstract}


\section{Introduction}

The study of noncommutative black holes is an active topic in both gravity
physics and modern geometry, see Ref. \cite{nic1} for a recent review of
results. Noncommutative geometry, quantum gravity and string/ brane theory
appear to be connected strongly in low energy limits. We can model physical
effects in such theories using deformations on noncommutative parameters of
some classes of exact and physically important solutions in general
relativity.

There were elaborated different approaches to quantum field theory
(including gauge and gravity models) on noncommutative spaces using, for
instance, the simplest example of a Moyal--Weyl spacetime, with and without
Seiberg--Witten maps and various applications in cosmology and black hole
physics, see \cite{asch1,szabo,mh,chaich1} and references therein. Our
constructions are based on the nonlinear connection formalism and Finsler
geometry methods in commutative and noncommutative geometry \cite{vsgg,vspdo}%
. They were applied to generalized Seiberg--Witten theories derived for the
Einstein gravity equivalently reformulated (at classical level, using
nonholonomic constraints) and/ or generalized as certain models of Poincar\'
e de Sitter gauge gravity \cite{vncgg}, see also extensions to nonholonomic
(super) gravity/string gravity theories and \cite{vcv}. Here we note that
there were also elaborated different models of noncommutative gauge gravity
theories \cite{moff,chams1,sah,card,nishino,ardalan,yang} based on
generalizations of some commutative/ complex/ nonsymmetric geometries. Our
approach was oriented to unify the constructions on commutative and
noncommutative gravity theories in the language of geometry of nonholonomic
manifolds/ bundle spaces\footnote{%
In modern geometry and applications to physics and mechanics, there are used
also equivalent terms like anholonomic and non--integrable manifolds; for
our purposes, it is convenient to use all such terms. A pair $(\mathbf{V},%
\mathcal{N})$, where $\mathbf{V}$ is a manifold and $\mathcal{N}$ is a
nonintegrable distribution on $\mathbf{V}$, is called a nonholonomic
manifold.}.

In Ref. \cite{vncg}, following the so--called anholonomic deformation method
(see recent reviews \cite{ijgmmp,vrflg}), we provided the first examples of
black hole/ ellipsoid/ toroidal solutions in noncommutative and/or
nonholonomic variables in Einstein gravity and gauge and string gravity
generalizations.\footnote{%
   The anholonomic deformation method should
be not confused with the Cartan's moving frame method even in the first case
"moving frames" can be also included. In our approach, we consider arbitrary
real/complex, in general, noncommutative/supersymmetric nonholonomic
distributions on certain manifolds and adapt the geometric constructions
with respect to such distributions. This results in (nonlinear) deformations
of connection and metric structures, which is not the case for moving
frames, when geometric objects are re--defined with respect to
moving/different systems of reference. Selecting some convenient nonholonomic distributions, we obtain separations of equations and
reparametrizations of variables in some physically important nonlinear
systems of partial differential equations which allows us to integrate such
systems in general forms. Then constraining correspondingly certain general
solutions, we select some subclasses of exact solutions, for instance, in general relativity.} The bulk of metrics for noncommutative black holes
reviewed in \cite{nic1} can be included as certain holonomic (non)
commutative configurations of nonholonomic solutions. This provides
additional arguments that a series of important physical effects for
noncommutative black holes can be derived by using nonholonomic and/ or
noncommutative deformations of well known solutions in general relativity.

In this article, we study two classes of Finsler type black hole solutions,
with zero and non--zero matter field sources/ cosmological constant, induced
by noncommutative anholonomic variables\footnote{%
let us say to be defined by certain quantum corrections in quasi--classical
limits of quantum gravity models} in Einstein gravity. Especially, we wish
to point out that such nonholonomic configurations may ''survive'' even in
the classical (commutative) limits and that Finsler type variables can be considered both in noncommutative gravity (defining complex nonholonomic distributions) and in Einstein gravity (stating some classes of real nonholonomic distributions).

Noncommutative relations on coordinates positively result in
generic off--diagonal metrics.\footnote{%
such metrics can not be diagonalized by coordinate transforms, see below
formulas (\ref{fuzcond}) and (\ref{dmnc})} The possibility to formulate a
geometric method of systematical derivation of exact solutions for
noncommutative deformations of general relativity brings a number of new
physical insights. For instance, we can generate new classes of noncommutative
Finsler like black hole solutions which are more general than the well known
Kerr solutions (when the off--diagonal terms of metric can be
modelled by rotation frames/ coordinates) and depend on noncommutative
parameters. Such black hole objects can be considered for models of Finsler
gravity on tangent bundles (with metrics and connections depending on
''velocities''), as nonholonomic configurations in Einstein gravity and
generalizations and/or extended to complex distributions in noncommutative
gravity. This allows us to examine important features  concerning deformations of horizons  and
topologies of locally anisotropic and/or noncommutative black holes, their
stability, phase structure and transitions, singularities and symmetries, quantum corrections,
 self--consistent imbedding into nontrivial solitonic
backgrounds etc.

In this context a systematical search of possible solutions in
noncommutative generalizations of Einstein gravity is of great significance. Using the fact that any type of noncommutative coordinate relations, and other structures like star product, noncommutative symmetries etc, can be considered as certain  complex distributions on a (pseudo) Riemannian manifold, or vector/tangent bundle, we can apply the formalism of
nonlinear connections and adapted geometric constructions
originally developed in Finsler and Lagrange geometry. The
resulting anholonomic deformation method provides us  not
only a new technique for finding solutions in certain gravity
theories but also  a promising unified geometric
scheme to (in general, nonholonomic) Ricci flow theory \cite%
{vspdo,vrf3,vrf4,vrf5}, deformation and A--brane quantization of gravity  \cite{vfqlf,vpla,avqg5,vbrane} and possible applications in modern particle physics.

The content of this work is as follows. In section \ref{s2} we outline the geometry of complex nonholonomic distributions defining noncommutative gravity models. Section \ref{s3} is devoted to a generalization of the
anholonomic frame method for constructing exact solutions with
noncommutative parameter. There are formulated the conditions when such solutions define effective off--diagonal metrics in Einstein gravity. We analyze noncommutative nonholonomic deformations of Schwarzschild spacetimes
in section \ref{s4} (being considered vacuum 
configurations, with nontrivial 
matter sources and with noncommutative ellipsoidal 
symmetries). In section %
\ref{s5} we provide a procedure of extracting black hole and rotoid configurations for small noncommutative parameters. We show how (non) commutative gravity models can be described using Finsler variables.
Finally, in section \ref{s6} there are formulated the conclusions of this work.

\section{Complex Nonholonomic Distributions and 
Noncommutative Gravity Models}

\label{s2} There exist many formulations of noncommutative geometry/gravity
based on nonlocal deformation of spacetime and field theories starting from
noncommutative relations of type
\begin{equation}
u^{\alpha }u^{\beta }-u^{\beta }u^{\alpha }=i\theta ^{\alpha \beta },
\label{fuzcond}
\end{equation}%
where $u^{\alpha }$ are local spacetime coordinates, $i$ is the imaginary
unity, $i^{2}=-1,$ and $\theta ^{\alpha \beta }$ is an anti--symmetric
second--rank tensor (which, for simplicity, for certain models, is taken to
be with constant coefficients). Following our unified approach to (pseudo)
Riemannian and Finsler--Lagrange spaces \cite{vrflg,vncg,vsgg} (using the
geometry of nonholonomic manifolds) we consider that for $\theta ^{\alpha
\beta }\rightarrow 0$ the local coordinates $u^{\alpha }$ are on a four
dimensional (4-d) nonholonomic manifold $\mathbf{V}$ of necessary smooth
class. Such spacetimes can be enabled with a conventional $2+2$ splitting
(defined by a nonholonomic, equivalently, anholonomic/ non--integrable real
distribution), when local coordinates $u=(x,y)$ on an open region $U\subset
\mathbf{V}$ are labelled in the form $u^{\alpha }=(x^{i},y^{a}),$ with
indices of type $i,j,k,...=1,2$ and $a,b,c...=3,4.$ The coefficients of
tensor like objects on $\mathbf{V}$ can be computed with respect to a
general (non--coordinate) local basis $e_{\alpha }=(e_{i},e_{a}).$\footnote{%
If $\mathbf{V}=TM$ is the total space of a tangent bundle $\left( TM,\pi
,M\right) $ on a two dimensional (2--d) base manifold $M,$ the values $x^{i}$
and $y^{a}$ are respectively the base coordinates (on a low--dimensional
space/ spacetime) and fiber coordinates (velocity like). Alternatively, we
can consider that $\mathbf{V}=V$ is a 4--d nonholonomic manifold (in
particular, a pseudo--Riemannian one) with local fibered structure.}

For our purposes, we consider a subclass of nonholonomic manifolds $\mathbf{%
V,}$ called N--anholononomic spaces (spacetimes, for corresponding
signatures), enabled with a nonintegrable distribution stating a
conventional horizontal (h) space, $\left( h\mathbf{V}\right) ,$ and
vertical (v) space, $\left( v\mathbf{V}\right) ,$
\begin{equation}
T\mathbf{V}=h\mathbf{V}\oplus v\mathbf{V}  \label{whitney}
\end{equation}%
which by definition determines a nonlinear connection (N--connection)
structure $\mathbf{N}=N_{i}^{a}(u)dx^{i}\otimes dy^{a},$ see details in \cite%
{ma1987,ma,vrflg,vncg,vsgg}. On a commutative $\mathbf{V,}$ any (prime)
metric $\mathbf{g=g}_{\alpha \beta }\mathbf{e}^{\alpha}\otimes \mathbf{e}%
^{\beta } $ (a Schwarzschild, ellipsoid, ring or other type solution, their
conformal transforms and nonholonomic deformations which, in general, are
not solutions of the Einstein equations) can be parametrized in the form
\begin{eqnarray}
\mathbf{g} &=&g_{i}(u)dx^{i}\otimes dx^{i}+h_{a}(u)\mathbf{e}^{a}\otimes
\mathbf{e}^{a},  \label{prime} \\
\mathbf{e}^{\alpha } &=&\mathbf{e}_{\ \underline{\alpha }}^{\alpha }(u)du^{%
\underline{\alpha }}=\left( e^{i}=dx^{i},\mathbf{e}%
^{a}=dy^{a}+N_{i}^{a}dx^{i}\right) .  \label{ddif}
\end{eqnarray}%
The nonholonomic frame structure is characterized by relations
\begin{equation}
\lbrack \mathbf{e}_{\alpha },\mathbf{e}_{\beta }]=\mathbf{e}_{\alpha }%
\mathbf{e}_{\beta }-\mathbf{e}_{\beta }\mathbf{e}_{\alpha }=w_{\alpha \beta
}^{\gamma }\mathbf{e}_{\gamma },  \label{anhrel}
\end{equation}%
where
\begin{equation}
\mathbf{e}_{\alpha }=\mathbf{e}_{\alpha \ }^{\ \underline{\alpha }%
}(u)\partial /\partial u^{\underline{\alpha }}=\left( \mathbf{e}%
_{i}=\partial /\partial x^{i}-N_{i}^{a}\partial /\partial
y^{a},e_{b}=\partial /\partial y^{b}\right)  \label{dder}
\end{equation}%
are dual to (\ref{ddif}). The nontrivial anholonomy coefficients are
determined by the N--connection coefficients $\mathbf{N}=\{N_{i}^{a}\}$
following formulas $w_{ia}^{b}=\partial _{a}N_{i}^{b}$ and $%
w_{ji}^{a}=\Omega _{ij}^{a},$ where
\begin{equation}
\Omega _{ij}^{a}=\mathbf{e}_{j}\left( N_{i}^{a}\right) -\mathbf{e}_{i}\left(
N_{j}^{a}\right)  \label{ncurv}
\end{equation}%
define the coefficients of N--connection curvature.\footnote{%
We use boldface symbols for spaces (and geometric objects on such spaces)
enabled with N--connection structure. Here we note that the particular
holonomic/ integrable case is selected by the integrability conditions $%
w_{\alpha \beta }^{\gamma }=0.$}

On a N--anholonomic manifold, it is convenient to work with the so--called
canonical distinguished connection (in brief, canonical d--connection $%
\widehat{\mathbf{D}}=\{\widehat{\mathbf{\Gamma }}_{\ \alpha \beta }^{\gamma
}\})$ which is metric compatible, $\widehat{\mathbf{D}}\mathbf{g}=0,$ and
completely defined by the coefficients of a metric $\mathbf{g}$ (\ref{prime}%
) and a N--connection $\mathbf{N,}$ subjected to the condition that the
so--called $h$-- and $v$--components of torsion are zero.\footnote{%
by definition, a d--connection is a linear connection preserving under
parallelism a given N--connection splitting (\ref{whitney}); in general, a
d--connection has a nontrivial torsion tensor but for the canonical
d--connection the torsion is induced by the anholonomy coefficients which in
their turn are defined by certain off--diagonal N--coefficients in the
corresponding metric} Using deformation of linear connections formula $\
\Gamma _{\ \alpha \beta }^{\gamma }=\widehat{\mathbf{\Gamma }}_{\ \alpha
\beta }^{\gamma }+\ Z_{\ \alpha \beta }^{\gamma },$ where $\nabla =\{\
\Gamma _{\ \alpha \beta }^{\gamma }\}$ is the Levi--Civita connection (this
connection is metric compatible, torsionless and completely defined by the
coefficients of the same metric structure $\mathbf{g),}$ we can perform all
geometric constructions in two equivalent forms: applying the covariant
derivative $\widehat{\mathbf{D}}$ and/or $\nabla .$ This is possible because
all values $\ \Gamma ,$ $\widehat{\mathbf{\Gamma }}$ and $\ Z$ are
completely determined in unique forms by $\mathbf{g}$ for a prescribed
nonholonomic splitting, see details and coefficient formulas in Refs. \cite%
{vfbh,vrflg,vncg,vsgg}.

Any class of noncommutative relations (\ref{fuzcond}) on a N--anholonomic
spacetime $\mathbf{V}$ defines additionally a complex distribution and
transforms this space into a complex nonholonomic manifold $\ ^{\theta }%
\mathbf{V.}$\footnote{%
Here we note that a noncommutative distribution of type (\ref{fuzcond})
mixes the h-- and v--components, for instance, of coordinates $x^{i}$ and $%
y^{a}.$ Nevertheless, it is possible to redefine the constructions in a
language of projective modules with certain conventional irreversible
splitting \ of type $T\ ^{\theta }\mathbf{V}=h\ ^{\theta }\mathbf{V}\oplus
v\ ^{\theta }\mathbf{V,}$ see details in \cite{vspdo} and Part III in \cite%
{vsgg}. Here we also note that we shall use the label $\theta $ both for
tensor like values $\theta _{\alpha \beta },$ or a set of parameters, for
instance, $\theta \delta _{\alpha \beta }.$} We shall follow the approach to
noncommutative geometry based on the Groenewold--Moyal product (star
product, or $\star $--product) \cite{gro,moyal} inspired by the foundations
of quantum mechanics \cite{weyl,wigner}. For the Einstein gravity and its
equivalent lifts on de Sitter/affine bundles and various types of
noncommutative Lagrange--Finsler geometries, we defined star products
adapted to N--connection structures \cite{vncgg,vcv,vncg}, see also \cite%
{vspdo} and Part III in \cite{vsgg} on alternative approaches with
nonholonomic Dirac operators and Ricci flows of noncommutative geometries.
In general, such constructions are related to deformations of the
commutative algebra of bounded (complex valued) continuous functions $%
\mathcal{C}(\mathbf{V})$ on $\mathbf{V}$ into a (noncommutative) algebra $\
^{\theta }\mathcal{A}(\mathbf{V}).$ There were considered different
constructions of $\ ^{\theta }\mathcal{A}$ corresponding to different
choices of the so--called ''symbols of operators'', see details and
references in \cite{asch1,szabo,weyl,wigner}, and the extended Weyl ordered
symbol $\mathcal{W},$ to get an algebra isomorphism with properties
\begin{equation*}
\mathcal{W}[\ ^{1}f\star \ ^{2}f]\equiv \mathcal{W}[\ ^{1}f]\mathcal{W}[\
^{2}f]=\ ^{1}\hat{f}\ \ ^{2}\hat{f},
\end{equation*}%
for $\ ^{1}f,\ ^{2}f\in \mathcal{C}(\mathbf{V})$ and $\ ^{1}\hat{f}\ ,\ ^{2}%
\hat{f}\in \ ^{\theta }\mathcal{A}(\mathbf{V}),$ when the induced $\star $%
--product is associative and noncommutative. Such a product can be
introduced on nonholonomic manifolds \cite{vncgg,vcv,vncg} using the
N--elongated partial derivatives (\ref{dder}),
\begin{equation}
\ ^{1}\hat{f}\star \ ^{2}\hat{f}=\sum\limits_{k=0}^{\infty }\frac{1}{k!}%
\left( \frac{i}{2}\right) ^{k}\theta ^{\alpha _{1}\beta _{1}}\ldots \theta
^{\alpha _{k}\beta _{k}}\mathbf{e}_{\alpha _{1}}\ldots \mathbf{e}_{\alpha
_{k}}\ ^{1}f(u)\ \mathbf{e}_{\beta _{1}}\ldots \mathbf{e}_{\beta _{k}}\
^{2}f(u).  \label{starpr}
\end{equation}%
For nonholonomic configurations, we have two types of \ ''noncommutativity''
given by relations (\ref{fuzcond}) and (\ref{anhrel}).

For a noncommutative nonholonomic spacetime model $\ ^{\theta }\mathbf{V}$
of a spacetime $\mathbf{V,}$ we can derive a N--adapted local frame
structure $\ \ ^{\theta }\mathbf{e}_{\alpha }=(\ \ ^{\theta }\mathbf{e}%
_{i},\ \ ^{\theta }\mathbf{e}_{a})$ which can be constructed by
noncommutative deformations of $\mathbf{e}_{\alpha },$%
\begin{eqnarray}
\ ^{\theta }\mathbf{e}_{\alpha \ }^{\ \underline{\alpha }} &=&\mathbf{e}%
_{\alpha \ }^{\ \underline{\alpha }}+i\theta ^{\alpha _{1}\beta _{1}}\mathbf{%
e}_{\alpha \ \alpha _{1}\beta _{1}}^{\ \underline{\alpha }}+\theta ^{\alpha
_{1}\beta _{1}}\theta ^{\alpha _{2}\beta _{2}}\mathbf{e}_{\alpha \ \alpha
_{1}\beta _{1}\alpha _{2}\beta _{2}}^{\ \underline{\alpha }}+\mathcal{O}%
(\theta ^{3}),  \label{ncfd} \\
\ ^{\theta }\mathbf{e}_{\ \star \underline{\alpha }}^{\alpha } &=&\mathbf{e}%
_{\ \underline{\alpha }}^{\alpha }+i\theta ^{\alpha _{1}\beta _{1}}\mathbf{e}%
_{\ \underline{\alpha }\alpha _{1}\beta _{1}}^{\alpha }+\theta ^{\alpha
_{1}\beta _{1}}\theta ^{\alpha _{2}\beta _{2}}\mathbf{e}_{\ \underline{%
\alpha }\alpha _{1}\beta _{1}\alpha _{2}\beta _{2}}^{\alpha }+\mathcal{O}%
(\theta ^{3}),  \notag
\end{eqnarray}%
subjected to the condition $\ ^{\theta }\mathbf{e}_{\ \star \underline{%
\alpha }}^{\alpha }\star \ ^{\theta }\mathbf{e}_{\alpha \ }^{\ \underline{%
\beta }}=\delta _{\underline{\alpha }}^{\ \underline{\beta }},$ for $\delta
_{\underline{\alpha }}^{\ \underline{\beta }}$ being the Kronecker tensor,
where $\mathbf{e}_{\alpha \ \alpha _{1}\beta _{1}}^{\ \underline{\alpha }}$
and $\mathbf{e}_{\alpha \ \alpha _{1}\beta _{1}\alpha _{2}\beta _{2}}^{\
\underline{\alpha }}$ can be written in terms of $\mathbf{e}_{\alpha \ }^{\
\underline{\alpha }},\theta ^{\alpha \beta }$ and the spin distinguished
connection corresponding to $\widehat{\mathbf{D}}.$ Such formulas were
introduced for noncommutative deformations of the Einstein and Sitter/
Poincar\'{e} like gauge gravity \cite{vncgg,vcv} and complex gauge gravity %
\cite{chams1} and then generalized for noncommutative nonholonomic
configurations in string/brane and generalized Finsler theories in Part III
in \cite{vsgg} and \cite{vncg,vspdo} (we note that we can also consider
alternative expansions in ''non'' N--adapted form working with the spin
connection corresponding to the Levi--Civita connection).

The noncommutative deformation of a metric (\ref{prime}), $\mathbf{g}$ $%
\rightarrow \ ^{\theta }\mathbf{g,}$ can be defined in the form%
\begin{equation}
\ ^{\theta }\mathbf{g}_{\alpha \beta }=\frac{1}{2}\eta _{\underline{\alpha }%
\underline{\beta }}\left[ \ ^{\theta }\mathbf{e}_{\alpha \ }^{\ \underline{%
\alpha }}\star \left( \ ^{\theta }\mathbf{e}_{\beta \ }^{\ \underline{\beta }%
}\right) ^{+}+\ ^{\theta }\mathbf{e}_{\beta \ }^{\ \underline{\beta }}\star
\left( \ ^{\theta }\mathbf{e}_{\alpha \ }^{\ \underline{\alpha }}\right) ^{+}%
\right] ,  \label{dmnc}
\end{equation}%
where $\left( \ldots \right) ^{+}$ denotes Hermitian conjugation and $\eta _{%
\underline{\alpha }\underline{\beta }}$ is the flat Minkowski space metric.
In N--adapted form, as nonholonomic deformations, such metrics were used for
constructing exact solutions in string/gauge/Einstein and Lagrange--Finsler
metric--affine and noncommutative gravity theories in Refs. \cite{vsgg,vncg}%
. In explicit form, formula (\ref{dmnc}) was introduced in \cite{chaich1}
for decompositions of type (\ref{ncfd}) performed for the spin connection
corresponding to the Levi--Civita connection. In our approach, the
''boldface'' formulas allow us to extend the formalism to various types of
commutative and noncommutative nonholonomic and generalized Finsler spaces
and to compute also noncommutative deforms of N--connection coefficients.

The target metrics resulting after noncommutative nonholonomic transforms
(to be investigated in this work) can \ be parametrized in general form
\begin{eqnarray}
\ ^{\theta }\mathbf{g} &=&\ ^{\theta }g_{i}(u,\theta )dx^{i}\otimes dx^{i}+\
^{\theta }h_{a}(u,\theta )\ ^{\theta }\mathbf{e}^{a}\otimes \ ^{\theta }%
\mathbf{e}^{a},  \label{target} \\
\ ^{\theta }\mathbf{e}^{\alpha } &=&\ ^{\theta }\mathbf{e}_{\ \underline{%
\alpha }}^{\alpha }(u,\theta )du^{\underline{\alpha }}=\left( e^{i}=dx^{i},\
^{\theta }\mathbf{e}^{a}=dy^{a}+\ ^{\theta }N_{i}^{a}(u,\theta
)dx^{i}\right) ,  \notag
\end{eqnarray}%
where it is convenient to consider conventional polarizations $\eta _{\ldots
}^{\ldots }$ when
\begin{equation}
\ ^{\theta }g_{i}=\check{\eta}_{i}(u,\theta )g_{i},\ \ ^{\theta }h_{a}=%
\check{\eta}_{a}(u,\theta )h_{a},\ ^{\theta }N_{i}^{a}(u,\theta )=\ \check{%
\eta}_{i}^{a}(u,\theta )N_{i}^{a},  \label{polf}
\end{equation}%
for $g_{i},h_{a},N_{i}^{a}$ given by a prime metric (\ref{prime}). How to
construct exact solutions of gravitational and matter field equations
defined by very general ansatz of type (\ref{target}), with coefficients
depending on arbitrary parameters $\theta $ and various types of integration
functions, in Einstein gravity and (non)commutative string/gauge/Finsler etc
like generalizations, is considered in \ Refs. \cite%
{ijgmmp,vsgg,vncg,vrflg,vfbh,avfbt}.\

In this work, we shall analyze noncommutative deformations induced by (\ref%
{fuzcond}) for a class of four dimensional (4--d (pseudo) Riemannian)
metrics (or 2--d (pseudo) Finsler metrics) defining (non) commutative
Finsler--Einstein spaces as exact solutions of the Einstein equations,
\begin{equation}
\ ^{\theta }\widehat{E}_{\ j}^{i}=\ _{h}^{\theta }\Upsilon (u)\delta _{\
j}^{i},\ \widehat{E}_{\ b}^{a}=\ _{v}^{\theta }\Upsilon (u)\delta _{\
b}^{a},\ ^{\theta }\widehat{E}_{ia}=\ \ ^{\theta }\widehat{E}_{ai}=0,
\label{eeqcdcc}
\end{equation}%
where $\ ^{\theta }\widehat{\mathbf{E}}_{\alpha \beta }=\{\ ^{\theta }%
\widehat{E}_{ij},\ ^{\theta }\widehat{E}_{ia},\ ^{\theta }\widehat{E}_{ai},\
^{\theta }\widehat{E}_{ab}\}$ are the components of the Einstein tensor
computed for the canonical distinguished connection (d--connection) $\ \
^{\theta }\widehat{\mathbf{D}},$ see details in \cite{vfbh,vrflg,ijgmmp,vsgg}
and, on Finsler models on tangent bundles, \cite{ma1987,ma}. Functions $\
_{h}^{\theta }\Upsilon $ and $\ \ _{v}^{\theta }\Upsilon $ are considered to
be defined by certain matter fields in a corresponding model of (non)
commutative gravity. The geometric objects in (\ref{eeqcdcc}) must be
computed using the $\star $--product (\ref{starpr}) and the coefficients
contain in general the complex unity $i.$ Nevertheless, it is possible to
prescribe such nonholonomic distributions on the ''prime'' $\mathbf{V}$
when, for instance,
\begin{equation*}
\ \widehat{E}_{\ j}^{i}(u)\rightarrow \ \widehat{E}_{\ j}^{i}(u,\theta ),\
_{h}^{\theta }\Upsilon (u)\rightarrow \ _{h}\Upsilon (u,\theta ),\ldots
\end{equation*}%
and we get generalized Lagrange--Finsler and/or (pseudo) Riemannian
geometries, and corresponding gravitational models, with parametric
dependencies of geometric objects on $\theta .$

Solutions of nonholonomic equations (\ref{eeqcdcc}) are typical ones for the
Finsler gravity with metric compatible d--connections\footnote{%
We emphasize that Finlser like coordinates can be considered on any
(pseudo), or complex Riemannian manifold and inversely, see discussions in %
\cite{vrflg,vfbh}. A real Finsler metric$\ \mathbf{f=\{f}$ $_{\alpha \beta
}\}$ can be parametrized in the canonical Sasaki form
\begin{equation*}
\ \mathbf{f}=\ f_{ij}dx^{i}\otimes dx^{j}+\ f_{ab}\ ^{c}\mathbf{e}%
^{a}\otimes \ ^{c}\mathbf{e}^{b},\ \ ^{c}\mathbf{e}^{a}=dy^{a}+\
^{c}N_{i}^{a}dx^{i},
\end{equation*}%
where the Finsler configuration is defied by 1) a fundamental real Finsler
(generating) function $F(u)=F(x,y)=F(x^{i},y^{a})>0$ if $y\neq 0$ and
homogeneous of type $F(x,\lambda y)=|\lambda |F(x,y),$ for any nonzero $%
\lambda \in \mathbb{R},$ with positively definite Hessian $\ f_{ab}=\frac{1}{%
2}\frac{\partial ^{2}F^{2}}{\partial y^{a}\partial y^{b}},$ when $\det |\
f_{ab}|\neq 0,$ see details in \cite{vfbh,vrflg}$.$ The \ Cartan canonical
N--connection structure $\ ^{c}\mathbf{N}=\{\ ^{c}N_{i}^{a}\}$ is defined
for an effective Lagrangian $L=F^{2}$ as $\ \ ^{c}N_{i}^{a}=\frac{\partial
G^{a}}{\partial y^{2+i}}$ \ with $G^{a}=\frac{1}{4}\ f^{a\ 2+i}\left( \frac{%
\partial ^{2}L}{\partial y^{2+i}\partial x^{k}}y^{2+k}-\frac{\partial L}{%
\partial x^{i}}\right) ,$ where $\ f^{ab}$ is inverse to $\ f_{ab}$ and
respective contractions of horizontal (h) and vertical (v) indices, $\
i,j,...$ and $a,b...,$ are performed following the rule: we can write, for
instance, an up $v$--index $a$ as $a=2+i$ and contract it with a low index $%
i=1,2.$ In brief, we shall write $y^{i}$ instead of $y^{2+i},$ or $y^{a}.$
Such formulas can be re--defined on complex manifolds/bundles for various
types of complex Finsler/Riemannian geometries/gravity models.} or in the
so--called Einsteing/string/brane/gauge gravity with nonholonomic/Finsler \
like variables. In the standard approach to the Einstein gravity, when $%
\widehat{\mathbf{D}}\rightarrow \nabla ,$ the Einstein spaces are defined by
metrics $\mathbf{g}$ as solutions of the equations
\begin{equation}
\ E_{\alpha \beta }=\Upsilon _{\alpha \beta },  \label{eeqlcc}
\end{equation}%
where $\ E_{\alpha \beta }$ is the Einstein tensor for $\nabla $ and $%
\Upsilon _{\alpha \beta }$ is proportional to the energy--momentum tensor of
matter in general relativity. Of course, for noncommutative gravity models
in (\ref{eeqlcc}), we must consider values of type $\ ^{\theta }\nabla ,\ \
^{\theta }E,\ \ ^{\theta }\Upsilon $ etc. Nevertheless, for certain general
classes of ansatz of primary metrics $\mathbf{g}$ on a $\mathbf{V}$ we can
reparametrize such a way the nonholonomic distributions on corresponding $\
^{\theta }\mathbf{V}$ that $\ ^{\theta }\mathbf{g}(u)=\mathbf{\tilde{g}}%
(u,\theta )$ are solutions of (\ref{eeqcdcc}) transformed into a system of
partial differential equations (with parametric dependence of coefficients
on $\theta )$ which after certain further restrictions on coefficients
determining the nonholonomic distribution can result in generic
off--diagonal solutions for general relativity.\footnote{%
the metrics for such spacetimes can not diagonalized by coordinate transforms%
}

\section{General Solutions with Noncommutative Parameters}

\label{s3} A noncommutative deformation of coordinates of type (\ref{fuzcond}%
) defined by $\theta $ together with correspondingly stated nonholonomic
distributions on $\ ^{\theta }\mathbf{V}$ transform prime metrics $\mathbf{g}
$ (for instance, a Schwarzschild solution on $\mathbf{V}$) into respective
classes of target metrics $\ ^{\theta }\mathbf{g}=\mathbf{\tilde{g}}$ as
solutions of Finsler type gravitational field equations (\ref{eeqcdcc})
and/or standard Einstein equations (\ref{eeqlcc}) in general gravity. The
goal of this section is to show how such solutions and their
noncommutative/nonholonomic transforms can be constructed in general form
for vacuum and non--vacuum locally anisotropic configurations.

We parametrize the noncommutative and nonholonomic transform of a metric $%
\mathbf{g}$ (\ref{prime}) into a $\ ^{\theta }\mathbf{g}=\mathbf{\tilde{g}}$
(\ref{target}) resulting from formulas (\ref{ncfd}), and (\ref{dmnc}) and
expressing of polarizations in (\ref{polf}), as $\check{\eta}_{\alpha
}(u,\theta )=\grave{\eta}_{\alpha }(u)+\mathring{\eta}_{\alpha }(u)\theta
^{2}+\mathcal{O}(\theta ^{4})$ in the form%
\begin{eqnarray}
\ ^{\theta }g_{i} &=&\grave{g}_{i}(u)+\mathring{g}_{i}(u)\theta ^{2}+%
\mathcal{O}(\theta ^{4}),\ ^{\theta }h_{a}=\grave{h}_{a}(u)+\mathring{h}%
_{a}(u)\theta ^{2}+\mathcal{O}(\theta ^{4}),  \notag \\
\ ^{\theta }N_{i}^{3} &=&\ ^{\theta }w_{i}(u,\theta ),\ \ ^{\theta
}N_{i}^{4}=\ ^{\theta }n_{i}(u,\theta ),  \label{coefm}
\end{eqnarray}%
where $\grave{g}_{i}=g_{i}$ and $\grave{h}_{a}=h_{a}$ for $\grave{\eta}%
_{\alpha }=1,$ but for general $\grave{\eta}_{\alpha }(u)$ we get
nonholonomic deformations which do not depend on $\theta .$

\subsection{Nonholonomic Einstein equations depending on noncommutative
parameter}

The gravitational field equations (\ref{eeqcdcc}) for a metric (\ref{target}%
) with coefficients (\ref{coefm}) and sources of type%
\begin{equation}
\ ^{\theta }\mathbf{\Upsilon }_{\beta }^{\alpha }=[\Upsilon
_{1}^{1}=\Upsilon _{2}(x^{i},v,\theta ),\Upsilon _{2}^{2}=\Upsilon
_{2}(x^{i},v,\theta ),\Upsilon _{3}^{3}=\Upsilon _{4}(x^{i},\theta
),\Upsilon _{4}^{4}=\Upsilon _{4}(x^{i},\theta )]  \label{sdiag}
\end{equation}%
transform into this system of partial differential equations\footnote{%
see similar details on computing the Ricci tensor coefficients $\ ^{\theta }%
\widehat{R}_{\beta }^{\alpha }$ for the canonical d--connection $\widehat{%
\mathbf{D}}$ in Parts II and III of \cite{vsgg} and reviews \cite%
{ijgmmp,vrflg}, revising those formulas for the case when the geometric
objects depend on noncommutative parameter $\theta $}:
\begin{eqnarray}
&&\ ^{\theta }\widehat{R}_{1}^{1}=\ ^{\theta }\widehat{R}_{2}^{2}=\frac{1}{%
2\ ^{\theta }g_{1}\ ^{\theta }g_{2}}\times  \label{ep1a} \\
&&\left[ \frac{\ ^{\theta }g_{1}^{\bullet }\ ^{\theta }g_{2}^{\bullet }}{2\
^{\theta }g_{1}}+\frac{(\ ^{\theta }g_{2}^{\bullet })^{2}}{2\ ^{\theta }g_{2}%
}-\ ^{\theta }g_{2}^{\bullet \bullet }+\frac{\ ^{\theta }g_{1}^{^{\prime }}\
^{\theta }g_{2}^{^{\prime }}}{2\ ^{\theta }g_{2}}+\frac{(\ ^{\theta
}g_{1}^{^{\prime }})^{2}}{2\ ^{\theta }g_{1}}-\ ^{\theta }g_{1}^{^{\prime
\prime }}\right] =-\Upsilon _{4}(x^{i},\theta ),  \notag \\
&&\ ^{\theta }\widehat{S}_{3}^{3}=\ ^{\theta }\widehat{S}_{4}^{4}=\frac{1}{%
2\ ^{\theta }h_{3}\ ^{\theta }h_{4}}\times  \label{ep2a} \\
&&\left[ \ ^{\theta }h_{4}^{\ast }\left( \ln \sqrt{|\ ^{\theta }h_{3}\
^{\theta }h_{4}|}\right) ^{\ast }-\ ^{\theta }h_{4}^{\ast \ast }\right]
=-\Upsilon _{2}(x^{i},v,\theta ),  \notag
\end{eqnarray}
\begin{eqnarray}
&&\ ^{\theta }\widehat{R}_{3i}=-\ ^{\theta }w_{i}\frac{\beta }{2\ ^{\theta
}h_{4}}-\frac{\alpha _{i}}{2\ ^{\theta }h_{4}}=0,  \label{ep3a} \\
&&\ ^{\theta }\widehat{R}_{4i}=-\frac{\ ^{\theta }h_{3}}{2\ ^{\theta }h_{4}}%
\left[ \ ^{\theta }n_{i}^{\ast \ast }+\gamma \ ^{\theta }n_{i}^{\ast }\right]
=0,  \label{ep4a}
\end{eqnarray}%
where, for $\ ^{\theta }h_{3,4}^{\ast }\neq 0,$%
\begin{eqnarray}
\alpha _{i} &=&\ ^{\theta }h_{4}^{\ast }\partial _{i}\phi ,\ \beta =\
^{\theta }h_{4}^{\ast }\ \phi ^{\ast },\ \gamma =\frac{3\ ^{\theta
}h_{4}^{\ast }}{2\ ^{\theta }h_{4}}-\frac{\ ^{\theta }h_{3}^{\ast }}{\
^{\theta }h_{3}},  \notag \\
\phi &=&\ln |\ ^{\theta }h_{3}^{\ast }/\sqrt{|\ ^{\theta }h_{3}\ ^{\theta
}h_{4}|}|,  \label{coefa}
\end{eqnarray}%
when the necessary partial derivatives are written in the form \ $a^{\bullet
}=\partial a/\partial x^{1},$ $a^{\prime }=\partial a/\partial x^{2},$\ $%
a^{\ast }=\partial a/\partial v.$ In the vacuum case, we must consider $%
\Upsilon _{2,4}=0.$ Various classes of (non) holonomic Einstein,
Finsler--Einstein and generalized spaces can be generated if the \ sources (%
\ref{sdiag}) are taken $\Upsilon _{2,4}=\lambda ,$ where $\lambda $ is a
nonzero cosmological constant, see examples of such solutons in Refs. \cite%
{vncg,vfbh,vrflg,avfbt,ijgmmp,vsgg}.

\subsection{Exact solutions for the canonical d--connection}

Let us express the coefficients of a target metric (\ref{target}), and
respective polarizations (\ref{polf}), in the form%
\begin{eqnarray}
\ ^{\theta }g_{k} &=&\epsilon _{k}e^{\psi (x^{i},\theta )},  \label{ansatz1}
\\
\ \ ^{\theta }h_{3} &=&\epsilon _{3}h_{0}^{2}(x^{i},\theta )\left[ f^{\ast
}\left( x^{i},v,\theta \right) \right] ^{2}|\varsigma \left( x^{i},v,\theta
\right) |\ ,  \notag \\
\ \ ^{\theta }h_{4} &=&\epsilon _{4}\left[ f\left( x^{i},v,\theta \right)
-f_{0}(x^{i},\theta )\right] ^{2}\ ,  \notag \\
\ ^{\theta }N_{k}^{3} &=&w_{k}\left( x^{i},v,\theta \right) ,\ ^{\theta
}N_{k}^{4}=n_{k}\left( x^{i},v,\theta \right) ,  \notag
\end{eqnarray}%
with arbitrary constants $\epsilon _{\alpha }=\pm 1,$ and $h_{3}^{\ast }\neq
0$ and $h_{4}^{\ast }\neq 0,$ when $f^{\ast }=0.$\footnote{%
The reason to chose such forms of parametrizations is that they generate
very general classes of exact solutions in general relativity and Finsler
gravity theories with one Killing vector symmetry, see details in \cite%
{vsgg,vrflg}. We proved in Ref. \cite{ijgmmp} that such solutions can be
generalized to possess dependencies on certain families of real parameters.
In this work, we modify the constructions for metrics when such $\theta$%
--parameters are for noncommutative deformations.}
By straightforward verification, or following methods outlined in Refs. \cite%
{ijgmmp,vsgg,vncg,vrflg}, we can prove that any off--diagonal metric
\begin{eqnarray}
&&\ \ _{\circ }^{\theta }\mathbf{g}=e^{\psi }\left[ \epsilon _{1}\
dx^{1}\otimes dx^{1}+\epsilon _{2}\ dx^{2}\otimes dx^{2}\right]  \notag \\
&&+\epsilon _{3}h_{0}^{2}\left[ f^{\ast }\right] ^{2}|\varsigma |\ \delta
v\otimes \delta v+\epsilon _{4}\left[ f-f_{0}\right] ^{2}\ \delta
y^{4}\otimes \delta y^{4},  \notag \\
&&\delta v=dv+w_{k}\left( x^{i},v,\theta \right) dx^{k},\ \delta
y^{4}=dy^{4}+n_{k}\left( x^{i},v,\theta \right) dx^{k},  \label{gensol1}
\end{eqnarray}%
defines an exact solution of the system of partial differential equations (%
\ref{ep1a})--(\ref{ep4a}), i.e. of the Einstein equation for the canonical
d--connection (\ref{eeqcdcc}) for a metric of type (\ref{target}) with the
coefficients of form (\ref{ansatz1}), if there are satisfied the conditions%
\footnote{%
we put the left symbol ''$\circ $'' in order to emphasize that such a metric
is a solution of gravitational field equations}:

\begin{enumerate}
\item function $\psi $ is a solution of equation $\epsilon _{1}\psi
^{\bullet \bullet }+\epsilon _{2}\psi ^{^{\prime \prime }}=\Upsilon _{4};$

\item the value $\varsigma $ is computed following formula
\begin{equation*}
\varsigma \left( x^{i},v,\theta \right) =\varsigma _{\lbrack 0]}\left(
x^{i},\theta \right) -\frac{\epsilon _{3}}{8}h_{0}^{2}(x^{i},\theta )\int
\Upsilon _{2}f^{\ast }\left[ f-f_{0}\right] dv
\end{equation*}%
and taken $\varsigma =1$ for $\Upsilon _{2}=0;$

\item for a given source $\Upsilon _{4},$ the N--connection coefficients are
computed following the formulas
\begin{eqnarray}
w_{i}\left( x^{k},v,\theta \right) &=&-\partial _{i}\varsigma /\varsigma
^{\ast },  \label{gensol1w} \\
n_{k}\left( x^{k},v,\theta \right) &=&\ ^{1}n_{k}\left( x^{i},\theta \right)
+\ ^{2}n_{k}\left( x^{i},\theta \right) \int \frac{\left[ f^{\ast }\right]
^{2}\varsigma dv}{\left[ f-f_{0}\right] ^{3}},  \label{gensol1n}
\end{eqnarray}%
and $w_{i}\left( x^{k},v,\theta \right) $ are arbitrary functions if $%
\varsigma =1$ for $\Upsilon _{2}=0.$
\end{enumerate}

It should be emphasized that such solutions depend on arbitrary nontrivial
functions $f$ (with $f^{\ast }\neq 0),$ $f_{0},$ $h_{0},$ $\ \varsigma
_{\lbrack 0]},$ $\ ^{1}n_{k}$ and $\ \ ^{2}n_{k},$ and sources $\Upsilon
_{2} $ and $\Upsilon _{4}.$ Such values for the corresponding
quasi--classical limits of solutions to metrics of signatures $\epsilon
_{\alpha }=\pm 1$ have to be defined by certain boundary conditions and
physical considerations.

Ansatz of type (\ref{target}) for coefficients (\ref{ansatz1}) with $%
h_{3}^{\ast }=0$ but $h_{4}^{\ast }\neq 0$ (or, inversely, $h_{3}^{\ast
}\neq 0$ but $h_{4}^{\ast }=0)$ consist more special cases and request a bit
different method of constructing exact solutions, see details in \cite{vsgg}.

\subsection{Off--diagonal solutions for the Levi--Civita connection}

The solutions for the gravitational field equations for the canonical
d--connection (which can be used for various models of noncommutative
Finsler gravity and generalizations) presented in the previous subsection
can be constrained additionally and transformed into solutions of the
Einstein equations for the Levi--Civita connection (\ref{eeqlcc}), all
depending, in general, on parameter $\theta .$ Such classes of metrics are
of type
\begin{eqnarray}
\ \ _{\circ }^{\theta }\mathbf{g} &=&e^{\psi (x^{i},\theta )}\left[ \epsilon
_{1}\ dx^{1}\otimes dx^{1}+\epsilon _{2}\ dx^{2}\otimes dx^{2}\right]
\label{eeqsol} \\
&&+h_{3}\left( x^{i},v,\theta \right) \ \delta v\otimes \delta v+h_{4}\left(
x^{i},v,\theta \right) \ \delta y^{4}\otimes \delta y^{4},  \notag \\
\delta v &=&dv+w_{1}\left( x^{i},v,\theta \right) dx^{1}+w_{2}\left(
x^{i},v,\theta \right) dx^{2},  \notag \\
\delta y^{4} &=&dy^{4}+n_{1}\left( x^{i},\theta \right) dx^{1}+n_{2}\left(
x^{i},\theta \right) dx^{2},  \notag
\end{eqnarray}%
with the coefficients restricted to satisfy the conditions
\begin{eqnarray}
\epsilon _{1}\psi ^{\bullet \bullet }+\epsilon _{2}\psi ^{^{\prime \prime }}
&=&\Upsilon _{4},\ h_{4}^{\ast }\phi /h_{3}h_{4}=\Upsilon _{2},  \label{ep2b}
\\
w_{1}^{\prime }-w_{2}^{\bullet }+w_{2}w_{1}^{\ast }-w_{1}w_{2}^{\ast }
&=&0,\ n_{1}^{\prime }-n_{2}^{\bullet }=0,  \notag
\end{eqnarray}%
for $w_{i}=\partial _{i}\phi /\phi ^{\ast },$ see (\ref{coefa}), for given
sources $\Upsilon _{4}(x^{k},\theta )$ and $\Upsilon _{2}(x^{k},v,\theta ).$
We note that the second equation in (\ref{ep2b}) relates two functions $%
h_{3} $ and $h_{4}$ and the third and forth equations from the mentioned
conditions select such nonholonomic configurations when the coefficients of
the canonical d--connection and the Levi--Civita connection are the same
with respect to N--adapted frames (\ref{ddif}) and (\ref{dder}), even such
connections (and corresponding derived Ricci and Riemannian curvature
tensors) are different by definition.

Even the ansatz (\ref{eeqsol}) depends on three coordinates $(x^{k},v)$ and
noncommutative parameter $\theta ,$ it allows us to construct more general
classes of solutions with dependence on four coordinates if such metrics can
be related by chains of nonholonomic transforms.

\section{Noncommutative Nonholonomic Deformations \newline
of Schwarz\-schild Metrics}

\label{s4}

Solutions of type (\ref{gensol1}) and/or (\ref{eeqsol}) are very general
ones induced by noncommutative nonholonomic distributions and it is not
clear what type of physical interpretation can be associated to such
metrics. In this section, we analyze certain classes of nonholonomic
constraints which allows us to construct black hole solutions and
noncommutative corrections to such solutions.

The goal of this subsection is to formulate the conditions when spherical
symmetric noncommutative (Schwarzschild type) configurations can be
extracted.

\subsection{Vacuum noncommutative nonholonomic configurations}

In the simplest case, we analyze a class of holonomic noncommutative
deformations, with $\ _{\shortmid }^{\theta }N_{i}^{a}=0,$\footnote{%
computed in Ref. \cite{chaich2}} of the Schwarzschild metric%
\begin{eqnarray*}
~^{Sch}\mathbf{g} &=&\ _{\shortmid }g_{1}dr\otimes dr+\ _{\shortmid }g_{2}\
d\vartheta \otimes d\vartheta +\ _{\shortmid }h_{3}\ d\varphi \otimes
d\varphi +\ _{\shortmid }h_{4}\ dt\otimes \ dt, \\
\ _{\shortmid }g_{1} &=&-\left( 1-\frac{\alpha }{r}\right) ^{-1},\ \
_{\shortmid }g_{2}=-r^{2},\ \ _{\shortmid }h_{3}=-r^{2}\sin ^{2}\vartheta ,\
\ _{\shortmid }h_{4}=1-\frac{\alpha }{r},
\end{eqnarray*}%
written in spherical coordinates $u^{\alpha }=(x^{1}=\xi ,x^{2}=\vartheta
,y^{3}=\varphi ,y^{4}=t)$ for $\alpha =2G\mu _{0}/c^{2},$ correspondingly
defined by the Newton constant $G,$ a point mass $\mu _{0}$ and light speed $%
c.$ Taking
\begin{eqnarray}
\ _{\shortmid }\grave{g}_{i} &=&\ _{\shortmid }g_{i},\grave{h}_{a}=\
_{\shortmid }h_{a},  \label{defaux} \\
\ _{\shortmid }\mathring{g}_{1} &=&-\frac{\alpha (4r-3\alpha )}{%
16r^{2}(r-\alpha )^{2}},\ _{\shortmid }\mathring{g}_{2}=-\frac{%
2r^{2}-17\alpha (r-\alpha )}{32r(r-\alpha )},  \notag \\
\ _{\shortmid }\mathring{h}_{3} &=&-\frac{(r^{2}+\alpha r-\alpha ^{2})\cos
\vartheta -\alpha (2r-\alpha )}{16r(r-\alpha )},\ _{\shortmid }\mathring{h}%
_{4}=-\frac{\alpha (8r-11\alpha )}{16r^{4}},  \notag
\end{eqnarray}%
for
\begin{equation*}
\ _{\shortmid }^{\theta }g_{i}=\ _{\shortmid }\grave{g}_{i}+\ _{\shortmid }%
\mathring{g}_{i}\theta ^{2}+\mathcal{O}(\theta ^{4}),\ \ \ _{\shortmid
}^{\theta }h_{a}=\ _{\shortmid }\grave{h}_{a}+\ _{\shortmid }\mathring{h}%
_{a}\theta ^{2}+\mathcal{O}(\theta ^{4}),
\end{equation*}%
we get a ''degenerated'' case of solutions (\ref{gensol1}), see details in
Refs. \cite{ijgmmp,vsgg,vncg,vrflg}, because $\ _{\shortmid }^{\theta
}h_{a}^{\ast }=\partial \ _{\shortmid }^{\theta }h_{a}/\partial \varphi =0$
which is related to the case of holonomic/ integrable off--diagonal metrics.
For such metrics, the deformations (\ref{defaux}) are just those presented
in Refs. \cite{chaich2,chaich1,nic1}.

A more general class of noncommutative deformations of the Schwarz\-schild
metric can be generated by nonholonomic transform of type (\ref{polf}) when
the metric coefficients polarizations, $\check{\eta}_{\alpha },$ and
N--connection coefficients, $\ \ _{\shortmid }^{\theta }N_{i}^{a},$ for
\begin{eqnarray*}
\ _{\shortparallel }^{\theta }g_{i} &=&\check{\eta}_{i}(r,\vartheta ,\theta
)\ _{\shortmid }g_{i},\ \ _{\shortparallel }^{\theta }h_{a}=\check{\eta}%
_{a}(r,\vartheta ,\varphi ,\theta )\ _{\shortmid }h_{a}, \\
\ \ _{\shortparallel }^{\theta }N_{i}^{3} &=&\ w_{i}(r,\vartheta ,\varphi
,\theta ),\ \ _{\shortparallel }^{\theta }N_{i}^{4}=\ n_{i}(r,\vartheta
,\varphi ,\theta ),
\end{eqnarray*}%
are constrained to define a metric (\ref{gensol1}) for $\Upsilon
_{4}=\Upsilon _{2}=0.$ The coefficients of such metrics, computed with
respect to N--adapted frames (\ref{ddif}) defined by $\ _{\shortparallel
}^{\theta }N_{i}^{a},$ can be re--parametrized in the form%
\begin{eqnarray}
\ \ \ _{\shortparallel }^{\theta }g_{k} &=&\epsilon _{k}e^{\psi (r,\vartheta
,\theta )}=\ _{\shortmid }\grave{g}_{k}+\delta \ _{\shortmid }\grave{g}%
_{k}+(\ _{\shortmid }\mathring{g}_{k}+\delta \ _{\shortmid }\mathring{g}%
_{k})\theta ^{2}+\mathcal{O}(\theta ^{4});  \label{ncfdm} \\
\ \ \ _{\shortparallel }^{\theta }h_{3} &=&\epsilon _{3}h_{0}^{2}\left[
f^{\ast }(r,\vartheta ,\varphi ,\theta )\right] ^{2} =  \notag \\
&& \left( \ _{\shortmid }\grave{h}_{3}+\delta \ _{\shortmid }\grave{h}%
_{3}\right) +\left( \ _{\shortmid }\mathring{h}_{3}+\delta \ _{\shortmid }%
\mathring{h}_{3}\right) \theta ^{2}+\mathcal{O}(\theta ^{4}),h_{0}=const\neq
0;  \notag \\
\ _{\shortparallel }^{\theta }h_{4} &=&\epsilon _{4}\left[ f(r,\vartheta
,\varphi ,\theta )-f_{0}(r,\vartheta ,\theta )\right] ^{2}=  \notag \\
&& \left(\ _{\shortmid }\grave{h}_{4}+\delta \ _{\shortmid }\grave{h}%
_{4}\right) +\left(\ _{\shortmid }\mathring{h}_{4}+\delta \ _{\shortmid }%
\mathring{h}_{4}\right) \theta ^{2}+\mathcal{O}(\theta ^{4}),  \notag
\end{eqnarray}%
where the nonholonomic deformations $\delta \ _{\shortmid }\grave{g}%
_{k},\delta \ _{\shortmid }\mathring{g}_{k},\delta \ _{\shortmid }\grave{h}%
_{a},\delta \ _{\shortmid }\mathring{h}_{a}$ are for correspondingly given
generating functions $\psi (r,\vartheta ,\theta )$ and $f(r,\vartheta
,\varphi ,\theta )$ expressed as series on $\theta ^{2k},$ for $k=1,2,... .$
Such coefficients define noncommutative Finsler type spacetimes being
solutions of the Einstein equations for the canonical d--connection. They
are determined by the (prime) Schwarzschild data $\ _{\shortmid }g_{i}$ and $%
\ _{\shortmid }h_{a}$ and certain classes on noncommutative nonholonomic
distributions defining off--diagonal gravitational interactions. In order to
get solutions for the Levi--Civita connection, we have to constrain (\ref%
{ncfdm}) additionally in a form to generate metrics of type (\ref{eeqsol})
with coefficients subjected to conditions (\ref{ep2b}) for zero sources $%
\Upsilon _{\alpha }.$

\subsection{Noncommutative deformations with nontrivial sources}

In the holonomic case, there are known such noncommutative generalizations
of the Schwarzschild metric (see, for instance, Ref. \cite{nic2,koba,rizzo}
and review \cite{nic1}) when%
\begin{eqnarray}
~^{ncS}\mathbf{g} &=&\ _{\intercal }g_{1}dr\otimes dr+\ _{\intercal }g_{2}\
d\vartheta \otimes d\vartheta +\ _{\intercal }h_{3}\ d\varphi \otimes
d\varphi +\ _{\intercal }h_{4}\ dt\otimes \ dt,  \notag \\
\ \ _{\intercal }g_{1} &=&-\left( 1-\frac{\ 4\mu _{0}\gamma }{\sqrt{\pi }r}%
\right) ^{-1},\ \ \ _{\intercal }g_{2}=-r^{2},  \label{ncsch} \\
\ \ \ _{\intercal }h_{3} &=&-r^{2}\sin ^{2}\vartheta ,\ \ \ _{\intercal
}h_{4}=1-\frac{\ 4\mu _{0}\gamma }{\sqrt{\pi }r},  \notag
\end{eqnarray}%
for $\gamma $ being the so--called lower incomplete Gamma function
\begin{equation*}
\gamma (\frac{3}{2},\frac{r^{2}}{4\theta })\doteqdot
\int\nolimits_{0}^{r^{2}}p^{1/2}e^{-p}dp,
\end{equation*}
is the solution of a noncommutative version of the Einstein equation
\begin{equation*}
\ \ \ ^{\theta }E_{\alpha \beta }=\frac{8\pi G}{c^{2}}\ \ ^{\theta
}T_{\alpha \beta },
\end{equation*}%
where $\ \ \ ^{\theta }E_{\alpha \beta }$ is formally left unchanged (i.e.
is for the commutative Levi--Civita connection in commutative coordinates)
but%
\begin{equation}
\ ^{\theta }T_{\ \beta }^{\alpha }=\left(
\begin{array}{cccc}
-p_{1} &  &  &  \\
& -p_{\perp } &  &  \\
&  & -p_{\perp } &  \\
&  &  & \rho _{\theta }%
\end{array}%
\right)  \label{ncs}
\end{equation}%
with $p_{1}=-\rho _{\theta }$ and $p_{\perp }=-\rho _{\theta }-\frac{r}{2}%
\partial _{r}\rho _{\theta }(r)$ is taken for a self--gravitating,
anisotropic fluid--type matter modeling noncommutativity.

Via nonholonomic deforms, we can generalize the solution (\ref{ncsch}) to
off--diagonal metrics of type
\begin{eqnarray}
\ ~_{\theta }^{ncS}\mathbf{g} &=&-e^{\psi (r,\vartheta ,\theta )}\left[ \
dr\otimes dr+d\vartheta \otimes d\vartheta \right]  \label{ncsolsch} \\
&&-h_{0}^{2}\left[ f^{\ast }(r,\vartheta ,\varphi ,\theta )\right]
^{2}|\varsigma (r,\vartheta ,\varphi ,\theta )|\ \delta \varphi \otimes
\delta \varphi  \notag \\
&&+\left[ f(r,\vartheta ,\varphi ,\theta )-f_{0}(r,\vartheta ,\theta )\right]
^{2}\ \delta t\otimes \delta t,  \notag \\
\delta \varphi &=&d\varphi +w_{1}(r,\vartheta ,\varphi ,\theta
)dr+w_{2}(r,\vartheta ,\varphi ,\theta )d\vartheta ,  \notag \\
\delta t &=&dt+n_{1}(r,\vartheta ,\varphi ,\theta )dr+n_{2}(r,\vartheta
,\varphi ,\theta )d\vartheta ,  \notag
\end{eqnarray}%
being exact solutions of the Einstein equation for the canonical
d--connection (\ref{eeqcdcc}) with locally anisotropically self--gravitating
source
\begin{equation*}
\ ^{\theta }\mathbf{\Upsilon }_{\beta }^{\alpha }=[\Upsilon
_{1}^{1}=\Upsilon _{2}^{2}=\Upsilon _{2}(r,\vartheta ,\varphi ,\theta
),\Upsilon _{3}^{3}=\Upsilon _{4}^{4}=\Upsilon _{4}(r,\vartheta ,\theta )].
\end{equation*}
Such sources should be taken with certain polarization coefficients when $%
\Upsilon \sim \eta T$ is constructed using the matter energy--momentum
tensor (\ref{ncs}).

The coefficients of metric (\ref{ncsolsch}) are computed to satisfy
correspondingly the conditions:

\begin{enumerate}
\item function $\psi (r,\vartheta ,\theta )$ is a solution of equation $\psi
^{\bullet \bullet }+\psi ^{^{\prime \prime }}=-\Upsilon _{4};$

\item for a nonzero constant $h_{0}^{2},$ and given $\Upsilon _{2},$
\begin{equation*}
\varsigma \left( r,\vartheta ,\varphi ,\theta \right) =\varsigma _{\lbrack
0]}\left( r,\vartheta ,\theta \right) +h_{0}^{2}\int \Upsilon _{2}f^{\ast }
\left[ f-f_{0}\right] d\varphi ;
\end{equation*}

\item the N--connection coefficients are
\begin{eqnarray*}
w_{i}\left( r,\vartheta ,\varphi ,\theta \right) &=&-\partial _{i}\varsigma
/\varsigma ^{\ast }, \\
n_{k}\left( r,\vartheta ,\varphi ,\theta \right) &=&\ ^{1}n_{k}\left(
r,\vartheta ,\theta \right) +\ ^{2}n_{k}\left( r,\vartheta ,\theta \right)
\int \frac{\left[ f^{\ast }\right] ^{2}\varsigma }{\left[ f-f_{0}\right] ^{3}%
}d\varphi .
\end{eqnarray*}
\end{enumerate}

The above presented class of metrics describes nonholonomic deformations of
the Schwarzschild metric into (pseudo) Finsler configurations induced by the
noncommutative parameter. Subjecting the coefficients of (\ref{ncsolsch}) to
additional constraints of type (\ref{ep2b}) with nonzero sources $\Upsilon
_{\alpha },$ we extract a subclass of solutions for noncommutative gravity
with effective Levi--Civita connection.

\subsection{Noncommutative ellipsoidal deformations}

In this section, we provide a method of extracting ellipsoidal
configurations from a general metric (\ref{ncsolsch}) with coefficients
constrained to generate solutions on the Einstein equations for the
canonical d--connection or Levi--Civita connection.

We consider a diagonal metric depending on noncommutative parameter $\theta$
(in general, such a metric is not a solution of any gravitational field
equations)
\begin{equation}
~^{\theta }\mathbf{g}=-d\xi \otimes d\xi -r^{2}(\xi )\ d\vartheta \otimes
d\vartheta -r^{2}(\xi )\sin ^{2}\vartheta \ d\varphi \otimes d\varphi
+\varpi ^{2}(\xi )\ dt\otimes \ dt,  \label{5aux1}
\end{equation}%
where the local coordinates and nontrivial metric coefficients are
parametriz\-ed in the form%
\begin{eqnarray}
x^{1} &=&\xi ,x^{2}=\vartheta ,y^{3}=\varphi ,y^{4}=t,  \label{5aux1p} \\
\check{g}_{1} &=&-1,\ \check{g}_{2}=-r^{2}(\xi ),\ \check{h}_{3}=-r^{2}(\xi
)\sin ^{2}\vartheta ,\ \check{h}_{4}=\varpi ^{2}(\xi ),  \notag
\end{eqnarray}%
for
\begin{equation*}
\xi =\int dr\ \left| 1-\frac{2\mu _{0}}{r}+\frac{\theta }{r^{2}}\right|
^{1/2}\mbox{\ and\ }\varpi ^{2}(r)=1-\frac{2\mu _{0}}{r}+\frac{\theta }{r^{2}%
}.
\end{equation*}%
For $\theta =0$ and variable $\xi (r),$ this metric is just the the
Schwarzschild solution written in spacetime spherical coordinates $%
(r,\vartheta ,\varphi ,t).$

Target metrics are generated by nonholonomic deforms with $g_{i}=\eta _{i}%
\check{g}_{i}$ and $h_{a}=\eta _{a}\check{h}_{a}$ and some nontrivial $%
w_{i},n_{i},$ where $(\check{g}_{i},\check{h}_{a})$ are given by data (\ref%
{5aux1p}) and parametrized by an ansatz of type (\ref{ncsolsch}),
\begin{eqnarray}
~_{\eta }^{\theta }\mathbf{g} &=&-\eta _{1}(\xi ,\vartheta ,\theta )d\xi
\otimes d\xi -\eta _{2}(\xi ,\vartheta ,\theta )r^{2}(\xi )\ d\vartheta
\otimes d\vartheta  \label{5sol1} \\
&&-\eta _{3}(\xi ,\vartheta ,\varphi ,\theta )r^{2}(\xi )\sin ^{2}\vartheta
\ \delta \varphi \otimes \delta \varphi +\eta _{4}(\xi ,\vartheta ,\varphi
,\theta )\varpi ^{2}(\xi )\ \delta t\otimes \delta t,  \notag \\
\delta \varphi &=&d\varphi +w_{1}(\xi ,\vartheta ,\varphi ,\theta )d\xi
+w_{2}(\xi ,\vartheta ,\varphi ,\theta )d\vartheta ,\   \notag \\
\delta t &=&dt+n_{1}(\xi ,\vartheta ,\theta )d\xi +n_{2}(\xi ,\vartheta
,\theta )d\vartheta ;  \notag
\end{eqnarray}
the coefficients of such metrics are constrained to be solutions of the
system of equations (\ref{ep1a})--(\ref{ep4a}).

The equation (\ref{ep2a}) for $\Upsilon _{2}=0$ states certain relations
between the coefficients of the vertical metric and respective polarization
functions,%
\begin{eqnarray}
h_{3} &=&-h_{0}^{2}(b^{\ast })^{2}=\eta _{3}(\xi ,\vartheta ,\varphi ,\theta
)r^{2}(\xi )\sin ^{2}\vartheta,  \label{aux41} \\
h_{4} &=&b^{2}=\eta _{4}(\xi ,\vartheta ,\varphi ,\theta )\varpi ^{2}(\xi ),
\notag
\end{eqnarray}%
for $|\eta _{3}|=(h_{0})^{2}|\check{h}_{4}/\check{h}_{3}|\left[ \left( \sqrt{%
|\eta _{4}|}\right) ^{\ast }\right] ^{2}.$ In these formulas, we have to
chose $h_{0}=const$ (it must be $h_{0}=2$ in order to satisfy the condition (%
\ref{ep2b})), where $\eta _{4}$ can be any function satisfying the condition
$\eta _{4}^{\ast }\neq 0.$ We generate a class of solutions for any function
$b(\xi ,\vartheta ,\varphi ,\theta )$ with $b^{\ast }\neq 0.$ For classes of
solutions with nontrivial sources, it is more convenient to work directly
with $\eta _{4},$ for $\eta _{4}^{\ast }\neq 0$ but, for vacuum
configurations, we can chose as a generating function, for instance, $h_{4},
$ for $h_{4}^{\ast }\neq 0.$

It is possible to compute the polarizations $\eta _{1}$ and $\eta _{2},$
when $\eta _{1}=\eta _{2}r^{2}=e^{\psi (\xi ,\vartheta )},$ from (\ref{ep1a}%
) with $\Upsilon _{4}=0,$ i.e. from $\psi ^{\bullet \bullet }+\psi ^{\prime
\prime }=0.$

Putting the above defined values of coefficients in the ansatz (\ref{5sol1}%
), we find a class of exact vacuum solutions of the Einstein equations
defining stationary nonholonomic deformations of the Sch\-warz\-schild
metric,
\begin{eqnarray}
~^{\varepsilon }\mathbf{g} &=&-e^{\psi (\xi ,\vartheta ,\theta )}\left( d\xi
\otimes d\xi +\ d\vartheta \otimes d\vartheta \right)  \label{5sol1a} \\
&&-4\left[ \left( \sqrt{|\eta _{4}(\xi ,\vartheta ,\varphi ,\theta )|}%
\right) ^{\ast }\right] ^{2}\varpi ^{2}(\xi )\ \delta \varphi \otimes \
\delta \varphi  \notag \\
&& +\eta _{4}(\xi ,\vartheta ,\varphi ,\theta )\varpi ^{2}(\xi )\ \delta
t\otimes \delta t,  \notag \\
\delta \varphi &=&d\varphi +w_{1}(\xi ,\vartheta ,\varphi ,\theta )d\xi
+w_{2}(\xi ,\vartheta ,\varphi ,\theta )d\vartheta ,\   \notag \\
\delta t &=&dt+\ ^{1}n_{1}(\xi ,\vartheta ,\theta )d\xi +\ ^{1}n_{2}(\xi
,\vartheta ,\theta )d\vartheta .  \notag
\end{eqnarray}%
The N--connection coefficients $w_{i}$ and $\ ^{1}n_{i}$ in (\ref{5sol1a})
must satisfy the last two conditions\ from (\ref{ep2b}) in order to get
vacuum metrics in Einstein gravity. Such vacuum solutions are for
nonholonomic deformations of a static black hole metric into (non) holonomic
noncommutative Einstein spaces with locally anistoropic backgrounds (on
coordinate $\varphi )$ defined by an arbitrary function $\eta _{4}(\xi
,\vartheta ,\varphi ,\theta )$ with $\partial _{\varphi }\eta _{4}\neq 0,$
an arbitrary $\psi (\xi ,\vartheta ,\theta )$ solving the 2--d Laplace
equation and certain integration functions $\ ^{1}w_{i}(\xi ,\vartheta
,\varphi ,\theta )$ and $\ ^{1}n_{i}(\xi ,\vartheta ,\theta ).$ The
nonholonomic structure of such spaces depends parametrically on
noncommutative parameter(s) $\theta .$

In general, the solutions from the target set of metrics (\ref{5sol1}), or (%
\ref{5sol1a}), do not define black holes and do not describe obvious
physical situations. Nevertheless, they preserve the singular character of
the coefficient $\varpi ^{2}(\xi )$ vanishing on the horizon of a
Schwarzschild black hole if we take only smooth integration functions for
some small noncommutative parameters $\theta .$ We can also consider a
prescribed physical situation when, for instance, $\eta _{4}$ mimics 3--d,
or 2--d, solitonic polarizations on coordinates $\xi ,\vartheta ,\varphi ,$
or on $\xi ,\varphi .$

\section{Extracting Black Hole and Rotoid Configurations}

\label{s5} From a class of metrics (\ref{5sol1a}) defining nonholonomic
noncommutative deformations of the Schwarzschild solution depending on
parameter $\theta ,$ it is possible to select locally anisotropic
configurations with possible physical interpretation of gravitational vacuum
configurations with spherical and/or rotoid (ellipsoid) symmetry.

\subsection{Linear parametric noncommutative polarizations}

Let us consider generating functions of type
\begin{equation}
b^{2}=q(\xi ,\vartheta ,\varphi )+ \bar{\theta} s(\xi ,\vartheta ,\varphi )
\label{gf1}
\end{equation}%
and, for simplicity, restrict our analysis only with linear decompositions
on a small dimensionless parameter $\bar{\theta}\sim \theta ,$ with $0<\bar{%
\theta}<<1.$ This way, we shall construct off--diagonal exact solutions of
the Einstein equations depending on $\ \bar{\theta} $ which for rotoid
configurations can be considered as a small eccentricity.\footnote{%
From a formal point of view, we can summarize on all orders $\ \left(\bar{%
\theta}\right) ^{2},$ $\left(\bar{\theta}\right) ^{3}...$ stating such
recurrent formulas for coefficients when get convergent series to some
functions depending both on spacetime coordinates and a parameter $\bar{%
\theta},$ see a detailed analysis in Ref. \cite{ijgmmp}.} For a value (\ref%
{gf1}), we get
\begin{equation*}
\left( b^{\ast }\right) ^{2}=\left[ (\sqrt{|q|})^{\ast }\right] ^{2}\left[ 1+%
\bar{\theta}\frac{1}{(\sqrt{|q|})^{\ast }}\left( \frac{s}{\sqrt{|q|}}\right)
^{\ast }\right]
\end{equation*}%
which allows us to compute the vertical coefficients of d--metric (\ref%
{5sol1a}) (i.e $h_{3}$ and $h_{4}$ and corresponding polarizations $\eta
_{3} $ and $\eta _{4})$ using formulas (\ref{aux41}).

On should emphasize that nonholonomic deformations are not obligatory
related to noncommutative ones. For instance, in a particular case, we can
generate nonholonomic deformations of the Schwarzschild solution not
depending on $\ \bar{\theta}:$ we have to put $\bar{\theta}=0$ in the above
formulas\ and consider $b^{2}=q$ and $\left( b^{\ast }\right) ^{2}=\left[ (%
\sqrt{|q|})^{\ast }\right] ^{2}.$ Such classes of black hole solutions are
analyzed in Ref. \cite{vfbh}.

Nonholonomic deformations to rotoid configurations can be generated for
\begin{equation}
q=1-\frac{2\mu (\xi ,\vartheta ,\varphi )}{r}\mbox{ and }s=\frac{q_{0}(r)}{%
4\mu ^{2}}\sin (\omega _{0}\varphi +\varphi _{0}),  \label{aux42}
\end{equation}%
with $\mu (\xi ,\vartheta ,\varphi )=\mu _{0}+\bar{\theta}\mu _{1}(\xi
,\vartheta ,\varphi )$ (locally anisotropically polarized mass) with certain
constants $\mu ,\omega _{0}$ and $\varphi _{0}$ and arbitrary
functions/polarizations $\mu _{1}(\xi ,\vartheta ,\varphi )$ and $q_{0}(r)$
to be determined from some boundary conditions, with $\ \bar{\theta}$
treated as the eccentricity of an ellipsoid.\footnote{%
we can relate $\bar{\theta}$ to an eccentricity because the coefficient $%
h_{4}=b^{2}=\eta _{4}(\xi ,\vartheta ,\varphi ,\ \bar{\theta} )$ $\varpi
^{2}(\xi )$ becomes zero for data (\ref{aux42}) if $r_{+}\simeq {2\mu _{0}}/[%
{1+\bar{\theta} \frac{q_{0}(r)}{4\mu ^{2}}\sin (\omega _{0}\varphi +\varphi
_{0})}],$ which is the ''parametric'' equation for an ellipse $r_{+}(\varphi
)$ for any fixed values $\frac{q_{0}(r)}{4\mu ^{2}},\omega _{0},\varphi _{0}$
and $\mu _{0}$} Such a noncommutative nonholonomic configuration determines
a small deformation of the Schwarzschild spherical horizon into an
ellipsoidal one (rotoid configuration with eccentricity $\bar{\theta}).$

We provide the general solution for noncommutative ellipsoidal black holes
determined by nonholonomic h--components of metric and N--connection
coefficients which ''survive'' in the limit $\ \bar{\theta}\rightarrow 0,$
i.e. such values do not depend on noncommutative parameter. Dependence \ on
noncommutativity is contained in v--components of metric. This class of
stationary rotoid type solutions is parametrized in the form
\begin{eqnarray}
~_{\theta }^{rot}\mathbf{g} &=&-e^{\psi }\left( d\xi \otimes d\xi +\
d\vartheta \otimes d\vartheta \right)  \notag \\
&&-4\left[ (\sqrt{|q|})^{\ast }\right] ^{2}\left[ 1+\bar{\theta}\frac{1}{(%
\sqrt{|q|})^{\ast }}\left( \frac{s}{\sqrt{|q|}}\right) ^{\ast }\right] \
\delta \varphi \otimes \ \delta \varphi  \notag \\
&&+\left( q+\bar{\theta}s\right) \ \delta t\otimes \delta t,  \label{rotoidm}
\\
\delta \varphi &=&d\varphi +w_{1}d\xi +w_{2}d\vartheta ,\ \delta t=dt+\
^{1}n_{1}d\xi +\ ^{1}n_{2}d\vartheta ,  \notag
\end{eqnarray}%
with functions $q(\xi ,\vartheta ,\varphi )$ and $s(\xi ,\vartheta ,\varphi
) $ given by formulas (\ref{aux42}) and N--connec\-ti\-on coefficients $%
w_{i}(\xi ,\vartheta ,\varphi )$ and $\ n_{i}=$ $\ ^{1}n_{i}(\xi ,\vartheta
) $ subjected to conditions
\begin{eqnarray*}
w_{1}w_{2}\left( \ln |\frac{w_{1}}{w_{2}}|\right) ^{\ast } &=&w_{2}^{\bullet
}-w_{1}^{\prime },\quad w_{i}^{\ast }\neq 0; \\
\mbox{ or \ }w_{2}^{\bullet }-w_{1}^{\prime } &=&0,\quad w_{i}^{\ast }=0;\
^{1}n_{1}^{\prime }(\xi ,\vartheta )-\ ^{1}n_{2}^{\bullet }(\xi ,\vartheta
)=0
\end{eqnarray*}%
and $\psi (\xi ,\vartheta )$ being any function for which $\psi ^{\bullet
\bullet }+\psi ^{\prime \prime }=0.$

For small eccentricities, a metric (\ref{rotoidm}) defines stationary
configurations for the so--called black ellipsoid solutions (their stability
and properties can be analyzed following the methods elaborated in \cite%
{vbe1,vbe2,vncg}, see also a summary of results and generalizations for
various types of locally anisotropic gravity models in Ref. \cite{vsgg}).
There is a substantial difference between solutions provided in this section
and similar black ellipsoid ones constructed in \cite{vfbh}. In this work,
such metrics transform into the usual Schwarzschild one if the values $%
e^{\psi },w_{i},\ ^{1}n_{i}$ have the corresponding limits for $\bar{\theta}%
\rightarrow 0,$ i.e. for commutative configurations. For ellipsoidal
configurations with generic off--diagonal terms, an eccentricity $%
\varepsilon $ may be non--trivial because of generic nonholonomic
constraints.

\subsection{Rotoids and noncommutative solitonic distributions}

There are static three dimensional solitonic distributions $\eta (\xi
,\vartheta ,\varphi ,\theta ),$ defined as solutions of a solitonic equation%
\footnote{%
a function $\eta $ can be a solution of any three dimensional solitonic and/
or other nonlinear wave equations}
\begin{equation*}
\eta ^{\bullet \bullet }+\epsilon (\eta ^{\prime }+6\eta \ \eta ^{\ast
}+\eta ^{\ast \ast \ast })^{\ast }=0,\ \epsilon =\pm 1,
\end{equation*}%
resulting in stationary black ellipsoid--solitonic noncommutative spacetimes
$^{\theta }\mathbf{V}$ \ generated as further deformations of a metric $%
~_{\theta }^{rot}\mathbf{g}$ (\ref{rotoidm}). Such metrics are of type
\begin{eqnarray}
~_{sol\theta }^{rot}\mathbf{g} &=&-e^{\psi }\left( d\xi \otimes d\xi +\
d\vartheta \otimes d\vartheta \right)  \label{solrot} \\
&&-4\left[ (\sqrt{|\eta q|})^{\ast }\right] ^{2}\left[ 1+\bar{\theta}\frac{1%
}{(\sqrt{|\eta q|})^{\ast }}\left( \frac{s}{\sqrt{|\eta q|}}\right) ^{\ast }%
\right] \ \delta \varphi \otimes \ \delta \varphi  \notag \\
&&+\eta \left( q+\bar{\theta}s\right) \ \delta t\otimes \delta t,  \notag \\
\delta \varphi &=&d\varphi +w_{1}d\xi +w_{2}d\vartheta ,\ \delta t=dt+\
^{1}n_{1}d\xi +\ ^{1}n_{2}d\vartheta ,  \notag
\end{eqnarray}%
where the N--connection coefficients are taken the same as for (\ref{rotoidm}%
).

For small values of $\bar{\theta},$ a possible spacetime noncommutativity
determines nonholonomic embedding of the Schwarzschild solution into a
solitonic vacuum. In the limit of small polarizations, when $|\eta |\sim 1,$
it is preserved the black hole character of metrics and the solitonic
distribution can be considered as on a Schwarzschild background. It is also
possible to take such parameters of $\eta $ when a black hole is
nonholonomically placed on a ''gravitational hill'' defined by a soliton
induced by spacetime noncommutativity.

A vacuum metric (\ref{solrot}) can be generalized for (pseudo) Finsler
spaces with canonical d--connection as a solution of equations $\widehat{%
\mathbf{R}}_{\alpha \beta }=0$ (\ref{eeqcdcc}) if the metric is generalized
to a subclass of (\ref{5sol1}) with stationary coefficients subjected to
conditions
\begin{eqnarray*}
&&\psi ^{\bullet \bullet }(\xi ,\vartheta ,\bar{\theta})+\psi ^{^{\prime
\prime }}(\xi ,\vartheta ,\bar{\theta})=0; \\
h_{3} &=&\pm e^{-2\ ^{0}\phi }\frac{\left( h_{4}^{\ast }\right) ^{2}}{h_{4}}%
\mbox{ for  given }h_{4}(\xi ,\vartheta ,\varphi ,\bar{\theta}),\ \phi =\
^{0}\phi =const; \\
w_{i} &=&w_{i}(\xi ,\vartheta ,\varphi ,\bar{\theta})%
\mbox{ are any
functions  }; \\
n_{i} &=&\ \ ^{1}n_{i}(\xi ,\vartheta ,\bar{\theta})+\ ^{2}n_{i}(\xi
,\vartheta ,\bar{\theta})\int \left( h_{4}^{\ast }\right)
^{2}|h_{4}|^{-5/2}dv,\ n_{i}^{\ast }\neq 0; \\
&=&\ ^{1}n_{i}(\xi ,\vartheta ,\bar{\theta}),n_{i}^{\ast }=0,
\end{eqnarray*}%
for $h_{4}=\eta (\xi ,\vartheta ,\varphi ,\bar{\theta})\left[ q(\xi
,\vartheta ,\varphi )+\bar{\theta}s(\xi ,\vartheta ,\varphi )\right] .$ In
the limit $\bar{\theta}\rightarrow 0,$ we get a Schwarzschild configuration
mapped nonholonomically on a N--anholonomic (pseudo) Riemannian spacetime
with a prescribed nontrivial N--connection structure.

The above constructed classes of noncommutative and/or nonholonomic black
hole type solutions (\ref{rotoidm}) and (\ref{solrot}) are stationary. It is
also possible to generalize such constructions for nonholonomic propagation
of black holes in extra dimension and/or as Ricci flows, in our case induced
by spacetime noncommutativity is also possible. We have to apply the
geometric methods elaborated in Refs. \cite{vs5dbh,vrf3,vrf4,vrf5}, see also
reviews of results, with solutions for the metric--affine gravity,
noncommutative generalizations etc, in \cite{vrflg,vsgg}.

\subsection{Noncommutative gravity and (pseudo) Finsler variables}

In Ref. \cite{vfbh}, we formulated a procedure of nonholonomic transforms of
(pseudo) Finsler metrics into (pseudo) Riemannian ones, and inversely, and
further deformations of both types of such metrics to exact solutions of the
Einstein equations. In this section, we show that such constructions can be
performed for nontrivial noncommutative parameters $\theta $ which emphasize
that (in general, complex) Finsler geometries can be induced by spacetime
noncommutativity. For certain types of nonholonomic distributions, the
constructions provide certain models of stationary black hole solutions. Of
course, such geometric/physical models are equivalent if they are performed
for the same canonical d--connection and/or Levi--Civita connection.

We summarize the main steps of such noncommutative complex Finsler --
(pseudo) Riemannian transform:

\begin{enumerate}
\item Let us consider a solution for (non)holonomic noncommutative
generalized Einstein gravity with a metric\footnote{%
we shall omit the left label $\theta $ in this section if this will not
result in ambiguities}
\begin{eqnarray*}
\ ^{\theta }\mathbf{\mathring{g}} &=&\mathring{g}_{i}dx^{i}\otimes dx^{i}+%
\mathring{h}_{a}(dy^{a}+\mathring{N}_{j}^{a}dx^{j})\otimes (dy^{a}+\mathring{%
N}_{i}^{a}dx^{i}) \\
&=&\mathring{g}_{i}e^{i}\otimes e^{i}+\mathring{h}_{a}\mathbf{\mathring{e}}%
^{a}\otimes \mathbf{\mathring{e}}^{a}=\mathring{g}_{i^{\prime \prime
}j^{\prime \prime }}e^{i^{\prime \prime }}\otimes e^{j^{\prime \prime }}+%
\mathring{h}_{a^{\prime \prime }b^{\prime \prime }}\mathbf{\mathring{e}}%
^{a^{\prime \prime }}\otimes \mathbf{\mathring{e}}^{b^{\prime \prime }}
\end{eqnarray*}%
related to an arbitrary (pseudo) Riemannian metric with transforms of type
\begin{equation}
\ ^{\theta }\mathbf{\mathring{g}}_{\alpha ^{\prime \prime }\beta ^{\prime
\prime }}=\ \mathbf{\mathring{e}}_{\ \alpha ^{\prime \prime }}^{\alpha
^{\prime }}\ \mathbf{\mathring{e}}_{\ \beta ^{\prime \prime }}^{\beta
^{\prime }}\ ^{\theta }\mathbf{g}_{\alpha ^{\prime }\beta ^{\prime }}
\label{secmap}
\end{equation}%
parametrized in the form%
\begin{equation*}
\mathring{g}_{i^{\prime \prime }j^{\prime \prime }}=g_{i^{\prime }j^{\prime
}}\mathbf{\mathring{e}}_{\ i^{\prime \prime }}^{i^{\prime }}\mathbf{%
\mathring{e}}_{\ j^{\prime \prime }}^{j^{\prime }}+h_{a^{\prime }b^{\prime }}%
\mathbf{\mathring{e}}_{\ i^{\prime \prime }}^{a^{\prime }}\mathbf{\mathring{e%
}}_{\ j^{\prime \prime }}^{b^{\prime }},\ \mathring{h}_{a^{\prime \prime
}b^{\prime \prime }}=g_{i^{\prime }j^{\prime }}\mathbf{\mathring{e}}_{\
a^{\prime \prime }}^{i^{\prime }}\mathbf{\mathring{e}}_{\ b^{\prime \prime
}}^{j^{\prime }}+h_{a^{\prime }b^{\prime }}\mathbf{\mathring{e}}_{\
a^{\prime \prime }}^{a^{\prime }}\mathbf{\mathring{e}}_{\ b^{\prime \prime
}}^{b^{\prime }}.
\end{equation*}
For $\mathbf{\mathring{e}}_{\ i^{\prime \prime }}^{i^{\prime }}=\delta _{\
i^{\prime \prime }}^{i^{\prime }},\mathbf{\mathring{e}}_{\ a^{\prime \prime
}}^{a^{\prime }}=\delta _{\ a^{\prime \prime }}^{a^{\prime }},$ we write (%
\ref{secmap}) as
\begin{equation*}
\mathring{g}_{i^{\prime \prime }}=g_{i^{\prime \prime }}+h_{a^{\prime
}}\left( \mathbf{\mathring{e}}_{\ i^{\prime \prime }}^{a^{\prime }}\right)
^{2},~\mathring{h}_{a^{\prime \prime }}=g_{i^{\prime }}\left( \mathbf{%
\mathring{e}}_{\ a^{\prime \prime }}^{i^{\prime }}\right) ^{2}+h_{a^{\prime
\prime }},
\end{equation*}%
i.e. in a form of four equations for eight unknown variables $\mathbf{%
\mathring{e}}_{\ i^{\prime \prime }}^{a^{\prime }}$ and $\mathbf{\mathring{e}%
}_{\ a^{\prime \prime }}^{i^{\prime }},$ and
\begin{equation*}
\ \mathring{N}_{i^{\prime \prime }}^{a^{\prime \prime }}=\mathbf{\mathring{e}%
}_{i^{\prime \prime }}^{\ i^{\prime }}\ \mathbf{\mathring{e}}_{\ a^{\prime
}}^{a^{\prime \prime }}\ N_{i^{\prime }}^{a^{\prime }}=N_{i^{\prime \prime
}}^{a^{\prime \prime }}.
\end{equation*}

\item We choose on $\ ^{\theta }\mathbf{V}$ a fundamental Finsler function
\begin{equation*}
F=\ ^{3}F(x^{i},v,\theta )+\ ^{4}F(x^{i},y,\theta )
\end{equation*}
inducing canonically a d--metric of type
\begin{eqnarray*}
\ ^{\theta }\mathbf{f} &=&\ f_{i}dx^{i}\otimes dx^{i}+\ f_{a}(dy^{a}+\
^{c}N_{j}^{a}dx^{j})\otimes (dy^{a}+\ ^{c}N_{i}^{a}dx^{i}), \\
&=&\ f_{i}e^{i}\otimes e^{i}+\ f_{a}\ ^{c}\mathbf{e}^{a}\otimes \ ^{c}%
\mathbf{e}^{a}
\end{eqnarray*}%
determined by data $\ \ ^{\theta }\mathbf{f}_{\alpha \beta }=\left[ \
f_{i},\ f_{a},\ ^{c}N_{j}^{a}\right] $ in a canonical N--elongated base $\
^{c}\mathbf{e}^{\alpha }=(dx^{i},\ ^{c}\mathbf{e}^{a}=dy^{a}+\
^{c}N_{i}^{a}dx^{i}).$

\item We define
\begin{equation*}
g_{i^{\prime }}=\ f_{i^{\prime }}\left( \frac{\mathring{w}_{i^{\prime }}}{\
^{c}w_{i^{\prime }}}\right) ^{2}\frac{h_{3^{\prime }}}{\ f_{3^{\prime }}}%
\mbox{ \ and \ }\ g_{i^{\prime }}=\ f_{i^{\prime }}\left( \frac{\mathring{n}%
_{i^{\prime }}}{\ ^{c}n_{i^{\prime }}}\right) ^{2}\frac{h_{4^{\prime }}}{\
f_{4^{\prime }}}.
\end{equation*}%
Both formulas are compatible if $\mathring{w}_{i^{\prime }}$ and $\mathring{n%
}_{i^{\prime }}$ are constrained to satisfy the conditions\footnote{%
see details in \cite{vfbh}}%
\begin{equation*}
\Theta _{1^{\prime }}=\Theta _{2^{\prime }}=\Theta ,
\end{equation*}%
where $\ \Theta _{i^{\prime }}=\left( \frac{\mathring{w}_{i^{\prime }}}{\
^{c}w_{i^{\prime }}}\right) ^{2}\left( \frac{\mathring{n}_{i^{\prime }}}{\
^{c}n_{i^{\prime }}}\right) ^{2},$ \ and $\ \Theta =\left( \frac{\mathring{w}%
_{1^{\prime }}}{\ ^{c}w_{1^{\prime }}}\right) ^{2}\left( \frac{\mathring{n}%
_{1^{\prime }}}{\ ^{c}n_{1^{\prime }}}\right) ^{2}=$\newline
$\left( \frac{\mathring{w}_{2^{\prime }}}{\ ^{c}w_{2^{\prime }}}\right)
^{2}\left( \frac{\mathring{n}_{2^{\prime }}}{\ ^{c}n_{2^{\prime }}}\right)
^{2}.$ \ Using $\Theta ,$ we compute
\begin{equation*}
g_{i^{\prime }}=\left( \frac{\mathring{w}_{i^{\prime }}}{\ ^{c}w_{i^{\prime
}}}\right) ^{2}\frac{\ f_{i^{\prime }}}{\ f_{3^{\prime }}}\mbox{\
and\ }h_{3^{\prime }}=h_{4^{\prime }}\Theta ,
\end{equation*}%
where (in this case) there is not summing on indices. So, we constructed the
data $g_{i^{\prime }},h_{a^{\prime }}$ and $w_{i^{\prime }},n_{j^{\prime }}.$

\item The values $\mathbf{\mathring{e}}_{\ i^{\prime \prime }}^{a^{\prime }}$
and $\mathbf{\mathring{e}}_{\ a^{\prime \prime }}^{i^{\prime }}$ are
determined as any nontrivial solutions of
\begin{equation*}
\mathring{g}_{i^{\prime \prime }}=g_{i^{\prime \prime }}+h_{a^{\prime
}}\left( \mathbf{\mathring{e}}_{\ i^{\prime \prime }}^{a^{\prime }}\right)
^{2},\ \mathring{h}_{a^{\prime \prime }}=g_{i^{\prime }}\left( \mathbf{%
\mathring{e}}_{\ a^{\prime \prime }}^{i^{\prime }}\right) ^{2}+h_{a^{\prime
\prime }},\ \mathring{N}_{i^{\prime \prime }}^{a^{\prime \prime
}}=N_{i^{\prime \prime }}^{a^{\prime \prime }}.
\end{equation*}%
For instance, we can choose
\begin{eqnarray*}
\mathbf{\mathring{e}}_{\ 1^{\prime \prime }}^{3^{\prime }} &=&\pm \sqrt{%
\left| \left( \mathring{g}_{1^{\prime \prime }}-g_{1^{\prime \prime
}}\right) /h_{3^{\prime }}\right| },\mathbf{\mathring{e}}_{\ 2^{\prime
\prime }}^{3^{\prime }}=0,\mathbf{\mathring{e}}_{\ i^{\prime \prime
}}^{4^{\prime }}=0 \\
\mathbf{\mathring{e}}_{\ a^{\prime \prime }}^{1^{\prime }} &=&0,\mathbf{%
\mathring{e}}_{\ 3^{\prime \prime }}^{2^{\prime }}=0,\mathbf{\mathring{e}}%
_{\ 4^{\prime \prime }}^{2^{\prime }}=\pm \sqrt{\left| \left( \mathring{h}%
_{4^{\prime \prime }}-h_{4^{\prime \prime }}\right) /g_{2^{\prime }}\right| }
\end{eqnarray*}%
and express
\begin{equation*}
e_{\ 1}^{1^{\prime }}=\pm \sqrt{\left| \frac{\ f_{1}}{g_{1^{\prime }}}%
\right| },\ e_{\ 2}^{2^{\prime }}=\pm \sqrt{\left| \frac{\ f_{2}}{%
g_{2^{\prime }}}\right| },\ e_{\ 3}^{3^{\prime }}=\pm \sqrt{\left| \frac{\
f_{3}}{h_{3^{\prime }}}\right| },\ e_{\ 4}^{4^{\prime }}=\pm \sqrt{\left|
\frac{\ f_{4}}{h_{4^{\prime }}}\right| }.
\end{equation*}
\end{enumerate}

Finally, in this seciton, we conclude that any model of noncommutative
nonhlonomic gravity with distributions of type (\ref{fuzcond}) and/or (\ref%
{whitney}) can be equivalently re--formulated as a Finsler gravity induced
by a generating function of type $F=\ ^{3}F+\ ^{4}F.$ In the limit $\theta
\rightarrow 0,$ for any solution $^{\theta }\mathbf{\mathring{g}},$ there is
a scheme of two nonholonomic transforms which allows us to rewrite the
Schwarzschild solution and its noncommutative/nonholonomic deformations as a
Finsler metric $\ ^{\theta }\mathbf{f.}$

\section{Concluding Remarks}

\label{s6} In this paper we have constructed new classes of exact solutions
with generic off--diagonal metrics depending on a noncommutative parameter $%
\theta .$ In particular we have studied nonholonomic noncommutative
deformations of Schwarzschild metrics which can induced by effective
energy--momentum tensors/ effective cosmological constants and/or
nonholonomic vacuum gravitational distributions. Such classes of solutions
define complex Finsler spacetimes, induced parametrically from Einstein
gravity, which can be equivalently modeled as complex Riemannian manifolds
enabled with nonholonomic distributions. We provided a procedure of
extracting stationary black hole configurations with ellipsoidal symmetry
and possible solitonic deformations.

In the presence of noncommutativity the nonholonomic frame structure and
matter energy--momentum tensor have contributions from the noncommutative
parameter. The anholonomic deformation method of constructing exact
solutions in gravity allows us to define real (pseudo) Finsler
configurations if we choose to work with the canonical distinguished
connection. Further restrictions on the metric and nonlinear connection
coefficients can be chosen in such a way that we can generate generic
off--diagonal solutions on general relativity.

Our geometric method allows us to consider immersing of different types of
(pseudo) Riemannian metrics, and/or exact solutions in Einstein gravity,
('prime' metrics) in noncommutative backgrounds which effectively polarize
the interaction constants, deforms nonholonomically the frame structure,
metrics and connections. The resulting 'target' metrics are positively
constructed to solve gravitational field equations but, in general, it is
difficult to understand what kind of physical importance they may have in
modern gravity. We have chosen small rotoid and solitonic noncommutative
deformations because there are explicit proofs that they are stable under
perturbations and have much similarity with stationary black hole solutions
in general relativity \cite{vbe1,vbe2,vncg}.

In this work, we emphasized constructions when black hole configurations are
imbedded self--consistently into nonholonomic backgrounds induced by
noncommutativity. The main difference from similar ellipsoidal
configurations and rotoid black holes considered in Ref. \cite{vfbh} is
that, in our case, the eccentricity is just a dimensionless variant of
noncommutative parameter (in general, we can construct solutions with an
infinite number of parameters of different origins, see details in \cite%
{ijgmmp}). So, such types of stationary black hole solutions are induced by
noncommutative deformations with additional nonholonomic constraints. They
are different from all those outlined in review \cite{nic1} and Refs. \cite%
{asch1,chaich1,chaich2,nic2,koba,rizzo} (those classes of noncommutative
solutions can be extracted from more general nonholonomic ones, constructed
in our works, as certain holonomic configurations).

Finally, we emphasize that the provided noncommutative generalization of the
anholonomic frame method can be applied to various types of commutative and
noncommutative (in general, nonsymmetric) models of gauge \cite{vncgg,vncg}
and string/brane gravity \cite{vbrane}, Ricci flows \cite%
{vspdo,vrf3,vrf4,vrf5} and nonholonomic quantum deformations of Einstein
gravity \cite{vfqlf,vpla,avqg5} as we emphasized in Refs. \cite%
{vsgg,vrflg,vcv}). All parameters of classical and quantum deformations
and/or of flow evolution, physical constants and coefficients of metrics and
connections, considered in those works, can be redefined to contain
effective noncommutative constants and polarizations.

\vskip5pt

\textbf{Acknowledgement: } S. V. is grateful to M. Anastasiei and G. Zet for
important discussions on nonholonomic geometry and noncommutative gauge
gravity and related exact solutions.

\end{document}